\pdfoutput=1
\documentclass[prd,eqsecnum,notitlepage,nofootinbib,preprintnumbers,tightenlines,longbibliography]{revtex4-1}

\usepackage{amsmath}
\usepackage{amssymb}
\usepackage{latexsym}
\usepackage{amsfonts}
\usepackage{slashed}
\usepackage{color}
\usepackage{bbold}

\usepackage{graphicx}% Include figure files
\usepackage{bm}% bold math
\usepackage{paralist}
\usepackage[colorlinks,linkcolor=blue,citecolor=blue]{hyperref}

\def\x{{\bm x}}
\def\y{{\bm y}}

\def\ls2{{\ell_s^2}}
\def\x{{\bm x}}
\def\dd{{\rm d}}

\newcommand\bga{\begin{align}}
\newcommand\nda{\end{align}}

\makeatletter
\newcommand{\pushright}[1]{\ifmeasuring@#1\else\omit\hfill$\displaystyle#1$\fi\ignorespaces}
\newcommand{\pushleft}[1]{\ifmeasuring@#1\else\omit$\displaystyle#1$\hfill\fi\ignorespaces}
\makeatother

\def\x{{\bm x}}

\def\y{{\bm y}}

\def\alphas{\alpha_{\rm s}}
\def\st{\begin{equation}}
\def\stp{\end{equation}}
\def\bg{\begin{eqnarray}}
\def\nd{\end{eqnarray}}
\def\Eq#1{eq.~(\ref{#1})}

\def\Fig#1{fig.~\ref{#1}}
\def\Sect#1{Section~\ref{#1}}
\def\Ref#1{ref.~\cite{#1}}

\def\llangle{\left\langle}
\def\rrangle{\right\rangle}

\def\LPM{Landau-Pomeranchuk-Migdal}

\def\nott#1{\setbox0=\hbox{$#1$}                % set a box for #1 
   \dimen0=\wd0                                 % and get its size
   \setbox1=\hbox{/} \dimen1=\wd1               % get size of /
   \ifdim\dimen0>\dimen1                        % #1 is bigger
      \rlap{\hbox to \dimen0{\hfil/\hfil}}      % so center / in box
      #1                                        % and print #1
   \else                                        % / is bigger
      \rlap{\hbox to \dimen1{\hfil$#1$\hfil}}   % so center #1
      /                                         % and print /
   \fi}                                         %

\def\st{\begin{equation}}
\def\stp{\end{equation}}
\def\bg{\begin{eqnarray}}
\def\nd{\end{eqnarray}}

\begin{document}

\title{A scaling relation between pA and AA collisions}

\author{G\"ok\c ce Ba\c sar}
\email{basar@tonic.physics.sunysb.edu}
\author{Derek Teaney}
\email{derek.teaney@stonybrook.edu}

\affiliation{Department of Physics and Astronomy,  Stony Brook University, Stony Brook, NY 11794, USA}

\begin{abstract}
We compare the flow-like correlations  in high multiplicity proton-nucleus
($p+A$) and nucleus-nucleus ($A+A$) collisions. At fixed multiplicity, the
correlations in these two colliding systems are strikingly similar, although
the system size is smaller in $p+A$.  Based on an independent cluster model and
a simple conformal scaling argument, where the ratio of the mean free path to
the system size stays constant at fixed multiplicity, we argue that flow in
$p+A$ emerges as a collective response to the fluctuations in the position of
clusters, just like in $A+A$ collisions. With several physically motivated and
parameter free rescalings of the recent LHC data, we show that this simple
model captures the essential physics of elliptic and triangular flow in $p+A$
collisions. 
%
%CHANGE
%
We also explore the implications of the model for jet  energy loss in
$p+A$, and predict slightly larger transverse momentum broadening 
in $p+A$ than in $A+A$ at the same multiplicity.
\end{abstract}

\maketitle

\date{\today}

\section{Introduction}
Recent measurements by the LHC  \cite{CMS:2012qk,Aad:2012gla,Abelev:2012ola}
and   RHIC \cite{Adare:2013piz} collaborations, have shown
that particle production in high multiplicity proton-nucleus ($p+A$) collisions  
exhibits striking long-range two-particle correlations. 
Indeed, the two-particle correlator in these high multiplicity 
events is qualitatively and even quantitatively similar to the corresponding  correlator in nucleus-nucleus ($A+A$) events. In the $A+A$ events the correlation
function has been successfully described with viscous hydrodynamics, where the
observed correlation arises from the collective response to the initial
geometry. 
The two particle angular correlation 
at large 
rapidity separation is decomposed into Fourier coefficients,
\st
\frac{\dd N_{\rm pairs} }{\dd \Delta \phi} = \frac{N_{\rm pairs} }{2\pi} \left[ 1 + 2\sum  V_{n\Delta}  \cos(n\Delta \phi)  \right] \, ,
\stp
and the Fourier coefficients  
are expressed in terms of the flow coefficients $v_n\{2\}$   
\st
v_n\{2\} \equiv \sqrt{ V_{n\Delta}  } \, .
\stp
The flow coefficients are measured as a function of momentum, particle type and  centrality and are compared to hydrodynamic simulations 
of the nucleus-nucleus event (see \Ref{Heinz:2013th} for an overview of this ongoing experimental and theoretical program).  

A comparison of the  flow coefficients in peripheral $A+A$ to high
multiplicity $p+A$ collisions, at the same overall multiplicity, shows that
the flow coefficients are similar in magnitude and depend on momentum in
similar  ways.  Indeed,
the two collision systems have the same 
integrated $v_{3}\{2\}$ to within 5\%. The striking
similarity between the observed correlations points to a common origin, and
challenges the hydrodynamic interpretation. Indeed, some
features of these correlations are reproduced by the Color Glass Condensate (CGC) without reference to the fluctuating geometry \cite{Dusling:2012cg,Dusling:2013oia}. However,
hydrodynamic simulations of $p+A$ events also qualitatively predicted
the correlations observed in the data \cite{Bozek:2011if,Bozek:2012gr,Bzdak:2013zma}, suggesting that
the origin of the flow in $p+A$ is similar to $A+A$. This has motivated
several phenomenological papers aiming to explain the observed
correlations and to differentiate these two approaches \cite{Bozek:2013sda,
Bzdak:2013rya,shuryak_zahed,McLerran:2013oju,Coleman-Smith:2013rla,Bjorken:2013boa,Dumitru:2013tja,LiYanPrivate}.

The purpose of the current paper is to give a concise explanation 
for the striking similarity of the flow harmonics in $p+A$ and $A+A$. We 
start by pointing out in \Sect{conformal} 
that 
%that provided the QCD dynamics
%is approximately conformal, then 
if the multiplicity is held fixed, and the initial
dynamics is approximately conformal, 
then the mean free path to system size is the same in the two colliding systems. 
The $p+A$ system is smaller than $A+A$, but hotter, and the  resulting 
response patterns in $p+A$ are scale-similar to the $A+A$ response.
Thus, it  is natural to expect that if a hydrodynamic response is supported in
$A+A$ collisions then a similar response is expected in high multiplicity $p+A$ collisions.
In $A+A$ collisions viscous corrections are rather  large in these  peripheral bins, and we expect similarly large corrections in $p+A$ collisions.

In \Sect{v2v3} we discuss elliptic and  triangular flow. After
scaling out the average geometry of the $A+A$ system (which can be 
done in a model independent way), we find that the integrated 
$v_{2}\{2\}$ in the two systems are essentially identical, 
as in the $v_3\{2\}$ case.
We point out that this is
not surprising in any picture based on an independent cluster model and
approximately conformal dynamics.  Since the process of scaling 
out the average geometry assumes that the observed $v_{2}\{2\}$ 
is a response to the geometry, the remarkable similarity of the
fluctuation driven 
$v_{2}\{2\}$ in the two systems strongly suggests that the response
in the $p+A$ system is also response to the geometry. The momentum 
dependence of the elliptic and triangular flow coefficients also supports
the conformal scaling outlined in \Sect{conformal}.
 
 Finally, in \Sect{jets} we discuss the implications of the conformal dynamics for jet energy loss in $p+A$, indicating a direction for future research.

\section{Conformal dynamics}
\label{conformal}

\subsection{$\ell_{\rm mfp}/L$ is the constant in $p+A$ and $A+A$ collisions at fixed multiplicity}

Working with a reasonable set of assumptions, we first note that the mean  free
path to system size is constant between high multiplicity $p+A$ and $A+A$
collisions, provided the multiplicity $\dd N/\dd y$ is kept fixed.  

Specifically, motivated by the Color Glass Condensate \cite{[{See for example, }]Gelis:2010nm}, we will 
adopt the following model for particle production in high multiplicity $p+A$ and  $A+A$ collisions.
\begin{enumerate}
   \item 
      First, we will assume a cluster 
model,  where the number of particles 
produced is proportional to the number of clusters.  
The typical momentum scale of the produced constituents in 
the initial state is set by the
number of clusters  per transverse area:
\st
Q_s^2 \sim  \frac{N_{\rm clust} }{\pi L^2 } \, ,
\stp
where $L$ is the transverse size of the high multiplicity events. 
We will assume that this is the only relevant momentum scale. 
A similar assumption was used in \Ref{McLerran:2013oju} to investigate the
systematics of particle spectra in high multiplicity $p+A$ collisions.

%
% CHANGE
%
\item  We will assume that the equilibration dynamics is conformal,
   so that the typical relaxation time $\tau_R$ is inversely proportional to 
   $Q_s$. Then, if $Q_s L$ is a sufficiently large number, 
   the system will equilibrate at a time $\tau_o$ with 
   $1/Q_s \ll \tau_o  \ll L$, and the initial temperature $T_{\rm o}$
   will be proportional to $Q_s$,\,  $T_{\rm o} \propto Q_s$.
   If the shear viscosity  
    is approximately conformal, $\eta \propto T^3$, then 
   viscous corrections due to transverse gradients 
   will be proportional to $1/(Q_s L)$. 
   %
   % CHANGE
   %
   Indeed, in  kinetic theory  
    transverse viscous corrections %arise 
  % from  deviations from equilibrium
  % distribution whose magnitude 
   are determined by the ratio of the mean free
   path  to the transverse size of the system. In conformal kinetics the
   initial mean free path is inversely proportional  
to $Q_s$, which is the only relevant momentum scale:
\st
\ell_{\rm mfp } \propto \frac{1}{Q_s} \, .
\stp
   
 %   estimate,     
%    where the viscous corrections to the stress-energy tensor are of the form 
%    $ \tau^{ij}\equiv-2 \eta \{ \partial^i u^j \}=\int {d^3 p\over (2\pi)^3\,p^0} p^i p^j \delta f(p)$.
%    Here $\delta f$ is the non-equilibrium correction to the particle distribution, which controls the viscous corrections to the flow, 
%     and $\{ \}$ denotes the symmetric, traceless component. The conformal approximation, then, leads 
%    to the order of magnitude estimation $\delta f \propto \tau_R \, \hat p^i \hat p^j\, \{\partial_i u_j \}\sim {1\over T L}\sim{1\over Q_s L}.$

\item Finally, we will also assume that the initial phase space distribution
   in a high multiplicity $p+A$ event is not parametrically different
   from a minimum bias event. For instance, an extremely high multiplicity
   di-jet event has a parametrically different initial phase space distribution.

\end{enumerate}

With these assumptions,
the multiplicity of a $p+A$ or $A+A$ event is 
\st
  \frac{\dd N}{\dd y}  \sim Q_s^2 L^2   \, . 
\stp

Then  mean free path to the transverse system size is  constant,
provided $\dd N/\dd y$ is kept fixed:
\st
\label{dNdy}
\frac{\ell_{\rm mfp}}{L}  \propto \frac{1}{Q_s L}  \propto \frac{1}{\sqrt{\dd N/\dd y} } \, .
\stp

This  line of reasoning  provides an extremely  simple explanation for why the
\emph{collective} response is similar in high multiplicity $p+A$ and peripheral $A+A$
collisions.  If the multiplicity is held fixed, then the 
conditions for the subsequent response in $p+A$ and $A+A$ are scale similar. The $p+A$ system is
smaller, but hotter, and  the initial temperature times the system size is
fixed.  If the subsequent expansion dynamics is approximately conformal,  then
the resulting collective response at a time, $\tau\, Q_s$, in the $p+A$ system
will be  equal to the $A+A$ response at the corresponding time.  We will adopt
this conformal scaling in what follows and investigate the attendant
consequences. 

%
%CHANGE
%
The preceding estimate for $\ell_{\rm mfp}/L$ in \Eq{dNdy} applies at
the earliest moments while the 
system is expanding longitudinally.
Specifically, we are considering times of order $\tau\sim \tau_o$ with  $Q_s \ll \tau_o \ll L$. A more relevant time 
scale for the development of elliptic flow is $\tau \sim L$. 
To estimate the size of $\ell_{\rm mfp}/L$ for $\tau \sim L$, we 
recall the Bjorken result for the decrease in the initial temperature 
due to the longitudinal expansion \cite{Bjorken:1982qr}
\st
T(\tau) = T_o \left(\frac{\tau_o}{\tau} \right)^{1/3} \,  ,
\stp
where $T_o$ and  $\tau_o$ scale with 
the saturation momentum,
$T_o \propto Q_s$ and  $\tau_o \propto Q_s^{-1}$.
Thus, at a time $\tau \sim L$ we have 
\st
\label{dNdy2}
\frac{\ell_{\rm mfp}}{L} \propto  \frac{1}{T(\tau) L }  \propto \frac{1}{(T_o L)^{2/3} } \propto 
\frac{1}{\sqrt[3]{\dd N/\dd y}  }.
\stp
This estimate shows that for an approximately conformal fluid, viscous corrections to elliptic flow scale as $(\dd N/\dd y)^{-1/3}$, and are again independent 
of the transverse size provided the multiplicity is held fixed.
 This is consistent with the findings of more complete hydrodynamic simulations,
 where the conformal assumptions of this section are only
 approximately respected.

%Clearly the freezeout temperature $T_{fo}$ constitutes an 
%additional scale in the problem which 
%influences several obersvables. Nevertheless if the freezout 

%that the dynamics of QCD is approximately conformal at  high enough energies,
%and that the only relevant momentum scale is set by the number of particles per
%area.  

%Both of these assumptions seem like a reasonable
%starting point for more sophisticated treatments. 

%If we assume that the response to the initial state follows an approximately conformal dynamics,
%%Since the
% \textcolor{blue}{some words on the freezout, Jean-Yves' work, etc..}

\section{Elliptic and Triangular Flow} 
\label{v2v3}

\subsection{Integrated flow coefficients}

Since the mean free path to system size is the same in the two colliding systems, we expect that the
integrated response $v_n /\epsilon_n$ should remain constant as  one changes from $p+A$ to $A+A$ collisions. 

We will adopt the independent cluster model 
to estimate $\epsilon_2\{2\}$ and $\epsilon_3\{2\}$ in $p+A$ and in $A+A$ \cite{Bhalerao:2006tp}.  Very recently, the independent cluster model has 
been used (independently) to estimate the fluctuations in $\epsilon_n$ in $p+A$ events \cite{Bzdak:2013rya,LiYanPrivate}.
In $A+A$, the independent cluster model quantitatively reproduces
the results of more sophisticated Glauber models~\cite{Bhalerao:2011bp}.
In the independent
cluster model, $N_{\rm clust}$ independent point like clusters are drawn from a  smooth
parent distribution, $\bar n(\x)$. As discussed in the previous section, the multiplicity of an event is proportional to the number of the clusters, and the fluctuations in the cluster density in the transverse plane, $n(\x)=\bar{n}(\x) + \delta n(\x)$,  source the anisotropic collective flow. These fluctuations are assumed to be random such that
\begin{equation}
\langle\delta n(\x)\delta n(\y)\rangle=\bar n(\x)\delta^{(2)}(\x-\y) \, .
\label{random}
\end{equation}
The angular brackets denote an average over events with a fixed number clusters.
We note that the current notation for the independent cluster model follows  \Ref{Jia:2012ju}. 

\subsubsection{Eccentricity and elliptic flow}
\label{eccentricity}

  The eccentricity is defined as:
    \begin{align}
   \epsilon_2 e^{i2\Phi_2}  \equiv&  \frac{\left\{ r^2 e^{i2\phi_s} \right\}  }{ \left\{r^2 \right\} },
 %  \epsilon_3 e^{i3\Phi_3}  \equiv&  \frac{\left\{ r^2 e^{i3\phi_s} \right\}  }{ \left\{r^2 \right\} },
\end{align}
where $\{ \ldots \}$ denotes an average over  the transverse plane in 
a single event. In $A+A$ collisions, there are two contributions to the
eccentricity. The first contribution is the average ellipticity of the overlap region in
non-central collisions. This contribution is parametrized by the standard eccentricity $\epsilon_s$, which is the eccentricity of the smooth parent
distribution. The second contribution comes from the fluctuations in the cluster density, which can be calculated using the statistics in \Eq{random}.
%In the independent cluster model 
%the root mean square fluctuations of the eccentricity 
%can  be calculated.
Using  eq.~(12) of 
\Ref{Bhalerao:2006tp} (see 
also eq.~(15) of \Ref{Jia:2012ju}),
the mean squared eccentricity  in $A+A$ collisions  is 
\begin{equation}
   (\epsilon_2\{2\})_{AA}^2  =
   \epsilon_s^2   +  \llangle \delta \epsilon_2^2 \rrangle +   
   \mathcal O\left(\frac{\epsilon_s^2 }{N_{\rm clust}} \right) + \mathcal O\left(\frac{1}{N_{\rm clust}^2 } \right). 
\end{equation}
where fluctuation driven eccentricity is
\begin{equation}
\langle\delta\epsilon_2^2\rangle=\frac{\llangle r^4 \rrangle }{N_{\rm clust} \llangle r^2 \rrangle^2 } \, .    
\label{r2}
\end{equation}
Here  the averages are over  the radial profile of the parent distribution, $\bar n(\x)$.
In $p+A$ collisions $\epsilon_s$ is presumably zero, and the squared eccentricity is determined only by  fluctuations:
\begin{equation}
 (\epsilon_2\{2\})_{pA}^2  = \llangle \delta \epsilon_2^2 \rrangle +   
   \mathcal O\left(\frac{1}{N_{\rm clust}^2} \right). 
\end{equation}  

\begin{figure}
   \includegraphics[width=0.52\textwidth]{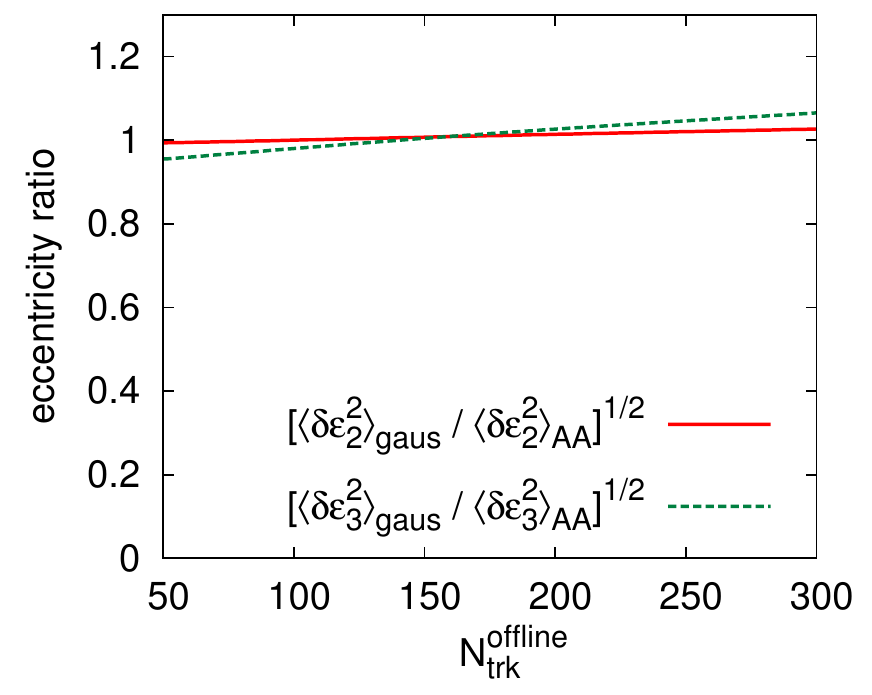}
   \caption{
      The ratio of fluctuation driven eccentricities,  $\delta \epsilon_2$ and
      $\delta \epsilon_3$, for a Gaussian profile compared to the Phobos Glauber Model~\cite{Alver:2008aq} as a function of $N_{\rm trk}^{\rm offline}$. The precise definition of these quantities are given in \Eq{e2ratio_eq} and \Eq{r3}.  
      \label{eratio_fig}
   }
\end{figure}

The
value of $\langle\delta\epsilon_2^2\rangle$ can differ in $p+A$ and $A+A$ collisions, since the spatial distribution of clusters  is not the same in the two systems.
However, we do not expect this difference to be very important in
determining $v_{2}\{2\}_{\rm pPb}/v_{2}\{2\}_{\rm PbPb}$,  since the relevant
parameter (at a fixed number of clusters) is the square root of a  geometric double ratio,
\st
\label{e2ratio_eq}
\sqrt{ \frac{\langle\delta\epsilon_2^2\rangle_{pA}}{\langle\delta\epsilon_2^2\rangle_{AA} } } = 
 \sqrt{ \frac{(\llangle r^4 \rrangle/\llangle r^2 \rrangle^2)_{pA} }{(\llangle r^4 \rrangle/\llangle r^2 \rrangle^2)_{AA} } } \, .
\stp
This parameter  will always be close to unity for any reasonable shape.
For example, 
comparing a hard sphere profile  $\bar n({\bm b})\propto \sqrt{1-b^2/R_0^2}$  
to   a Gaussian,  one finds
\st
\sqrt{ \frac{\langle\delta\epsilon_2^2\rangle_{\rm hard-sphere}}{\langle\delta\epsilon_2^2\rangle_{\rm Gaussian} } } 
\approx 0.85 \, .
\stp
Thus, even with  rather different profiles, the difference in 
the fluctuation-driven eccentricities  $\sqrt{\llangle \delta \epsilon_2^2 \rrangle }$ is only 15\%. 
More importantly, demanding similar eccentricities to 5\% accuracy does not 
require a fine tuning. Since a Gaussian profile for the $p+A$
event would arise in any diffusive process, this profile seems particularly
important.
In \Fig{eratio_fig} we have computed the ratio in \Eq{e2ratio_eq} for a Gaussian
profile and the Phobos Glauber Model as a function of $N_{\rm trk}^{\rm offline}$,
and the result is unity to a few percent accuracy.
The relation between $N_{\rm trk}^{\rm offline}$  and $N_{\rm part}$ is
from \Ref{CMS_flow}.
%the profiles do not need to be fine-tuned
%to have  the same $\sqrt{\llangle \delta \epsilon_2^2 \rrangle}$ to 
%within a few percent.

In the framework of linear response, elliptic flow is understood as a
collective response  to the eccentricity of the initial geometry
such that $v_2=k_2\epsilon_2$. The linear response coefficient
$k_2$ depends only on the ratio of the mean free path to the system size.
Therefore, the conformal scaling of \Sect{conformal} predicts that 
the $k_2$ coefficient is the same
in $p+A$ and $A+A$ collisions at fixed multiplicity. In order to fairly compare the $v_2$
in $p+A$ and $A+A$ we should first remove dependence on the average geometry,
and isolate the fluctuation-driven  $v_2$ in $A+A$.
 This can be achieved by scaling the $v_2$ in $A+A$  by the appropriate factor 
 $\sqrt{\epsilon_2\{2\}^2 - \epsilon_s^2}/{\epsilon_2\{2\} }$, so that
 \begin{align}
    \frac{\sqrt{\epsilon_2\{2\}^2 - \epsilon_s^2}}{\epsilon_2\{2\} }   (v_2\{2\})_{AA} =& k_2 \sqrt{\langle\delta\epsilon_2^2\rangle_{AA}} \, , \\
    (v_2\{2\})_{pA} =& k_2 \sqrt{\langle\delta\epsilon_2^2\rangle_{pA}} \, .
     \end{align}
 It is useful to define a rescaled $v_2\{2\}$ for $A+A$ that isolates
 the fluctuations
 %the fluctuating component
 %where the effect of the average ellipticity of the smooth parent distribution is taken out by a rescaling 
 \begin{align}
   (v_2\{2\})_{\rm PbPb, rscl}\equiv\sqrt{1-{ \epsilon_s^2\over\epsilon_2\{2\}^2} }\, (v_2\{2\})_{\rm PbPb}  \, .
   \label{v2_scaling}
 \end{align}
 We calculated the rescaling factor in \Eq{v2_scaling} with the Phobos Glauber
 Model~\cite{Alver:2008aq} using the relation between $N_{\rm trk}^{\rm offline}$
 and centrality provided by the CMS collaboration \cite{CMS_flow}.  It should be stressed
 that this rescaling factor
 is a non-trivial
 function of impact parameter and multiplicity,  and that there are no free
 parameters. This factor  is completely
 determined by the Glauber model simulation of the $A+A$ event. 
 %
 % CHANGES
 %
Similar rescalings have been used to explain the difference between
$v_2\{2\}$ and $v_2\{4\}$  as a function of centrality in A+A collisions
\cite{Bhalerao:2006tp,Aad:2013xma}.  Indeed, as in the current analysis, this difference primarily reflects the relative size of the average and fluctuating eccentricities \cite{Bhalerao:2006tp}.

\begin{figure}
\includegraphics[width=0.9\textwidth]{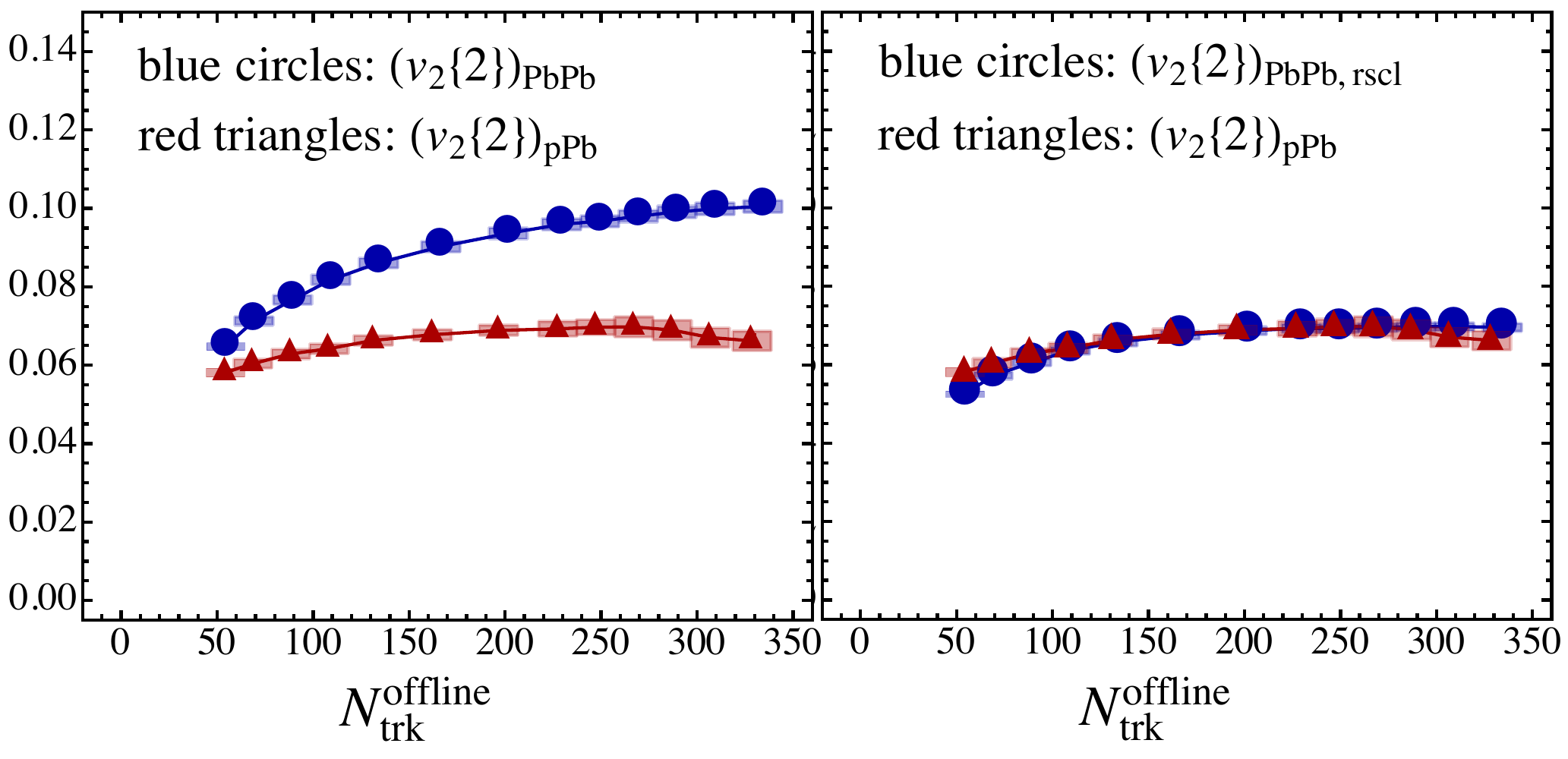}
\caption{The integrated $v_2\{2\}$ for PbPb and pPb vs. multiplicity from \cite{CMS_flow}. Left: Original values.  Right: The fluctuation dependent elliptic flow, $(v_2\{2\})_{\rm PbPb,  rscl}=\sqrt{1-\epsilon_s^2/\epsilon_2\{2\}^2} \; (v_2\{2\})_{\rm PbPb}$, compared to $(v_2\{2\})_{\rm pPb}$. The scaling factor is extracted using the Phobos Glauber Model \cite{Alver:2008aq} in $A+A$ simulations, and is not a fit. }
\label{fig_int_v22}
\end{figure}

 In \Fig{fig_int_v22} we compare the fluctuation driven part of the
 $(v_2\{2\})_{\rm PbPb}$ to $(v_2\{2\})_{\rm pPb}$. The data is taken from the CMS
 collaboration \cite{CMS_flow}. The striking agreement between these curves
 after this geometric rescaling  is a strong indication that the elliptic flow
 in $p+A$ stems from the same collective physics that determines
 the elliptic flow in $A+A$.  As this rescaling was motivated
 by geometry, the response in the $p+A$ system should also be driven
 by the fluctuating geometry.  
 Furthermore, the assumption that 
 the two systems are related by a conformal rescaling, where the linear response
 coefficients are  the same at fixed multiplicity, provides 
 a concise explanation for the similar $v_2\{2\}$ in the two systems.

 %
 % CHANGES
 % 
It is worth emphasizing  that to calculate the eccentricity correction
factor, $\sqrt{1 -\epsilon_s^2/\epsilon_2\{2\}^2}$, we  are using the $A+A$
Glauber model and not the $p+A$ Glauber model.  There are significant
uncertainties even in the $A+A$ Glauber model for these peripheral bins. However,
these uncertainties correct a relatively modest correction factor, and are
therefore small in Fig. 2. The uncertainty in $k_2 = v_2\{2\}/\epsilon_2\{2\}$
is larger (see Fig. 6 of \Ref{Alver:2008zza}), but the precise value of $k_2$
is not needed for this analysis.

% It is also worth to mention
% that, both the deviation from the conformality and the difference in the
% eccentricity fluctuations for $p+A$ and $A+A$ has to be negligibly small for
% the agreement, which is seen in the data,  to hold.

%
%We computed computed with the Phobos Glauber model. 
%Then we find  a simple scaling prediction
%\st
%\frac{\left( v_2\{2\} /v_3\{2\} \right)_{pA} }{  
%\left( \sqrt{1 - \epsilon_s^2/\epsilon_2\{2\}^2 } \;  v_2\{2\} / v_3\{2\} \right)_{\rm PbPb} } 
%   = 1
%   \label{double_ratio}
%\stp
%A plot of the experimental value  this double ratio is shown in \Fig{fig_double_ratio}.
%\begin{figure}[h]
%\includegraphics[scale=0.8]{double_ratio.pdf}
%\caption{The double ratio (\ref{double_ratio}) which is predicted to be unity according to the independent cluster model and conformal dynamics. The dashed line is 1/0.96 which is associated with the empirical ratio $(v_3\{2\})_{\rm pPb}/(v_3\{2\})_{PbPb}$.}
%\label{fig_double_ratio}
%\end{figure}
%
%To be able to predict the absolute value of $(v_2\{2\})_{pA}$   
%we need to know the square root geometric double ratio
%\st
%R  =   \sqrt{ \frac{
%\left(\llangle r^4 \rrangle/\llangle r^2 \rrangle^2 \right)_{pA}  
%} 
%%
%{  \left( \llangle r^4 \rrangle /\llangle r^2 \rrangle^2 \right)_{\rm PbPb}      } }
%\stp

\subsubsection{Triangularity and triangular flow}

Similar observations hold for $v_3\{2\}$. Since the triangularity is  produced by the fluctuations in the cluster density and not the average geometry, the comparison is more direct. We define the triangularity
%with an 
%$r^2$ weight
\begin{equation}
   \label{r3weight}
   \epsilon_3 e^{i3\Phi_3}\equiv{\{r^3 e^{i3\phi_s}\}\over\{r^2\}^{3/2}} \, ,
\end{equation}
 and compute the squared fluctuations of $\epsilon_3$ in $p+A$ and $A+A$
 in  the independent cluster model~\cite{Bhalerao:2011bp,Jia:2012ju}
\begin{equation}
\langle\delta\epsilon_3^2\rangle={\langle r^6\rangle\over N_{\rm clust}\langle r^2\rangle^3} \, .
\label{r3}
\end{equation}
We have used an $r^3$ weight to define the triangularity. 
If an $r^2$ weight is used, 
all fluctuation-driven eccentricities are equal~\cite{Jia:2012ju}, $i.e.$
   \st
   \llangle \delta \epsilon_2^2 \rrangle = \llangle \delta \epsilon_3^2 \rrangle  = \frac{\llangle r^4 \rrangle}{N_{\rm clust} \llangle r^2 \rrangle^2 } \,   \qquad\qquad (\mbox{$r^2$ weight}) \, .
   \stp
The optimal radial weight should be chosen to maximize the 
correlation between the flow response and the geometric predictor~\cite{Gardim:2011xv}.
With either weight, the relevant parameter for determining the ratio of $v_3$ in the two colliding systems is 
\st
\sqrt{ \frac{\langle\delta\epsilon_3^2\rangle_{pA}}{\langle\delta\epsilon_3^2\rangle_{AA} } } \, . 
 % = \sqrt{ \frac{(\llangle r^6 \rrangle/\llangle r^2 \rrangle^3)_{pA} }{(\llangle r^6 \rrangle/\llangle r^2 \rrangle^3)_{AA} } } \, .
\stp
This will be close to unity for reasonable profiles, 
though the deviation from unity is potentially larger when the $r^3$ weight is
used. For a Gaussian profile $p+A$ profile (which seems particularly well motivated), 
we compare $\llangle \delta \epsilon_3^2 \rrangle_{\rm gaus}$
to the nuclear profile in \Fig{eratio_fig} and the result is unity to within
5\%.

\begin{figure}
\includegraphics[width=0.47\textwidth]{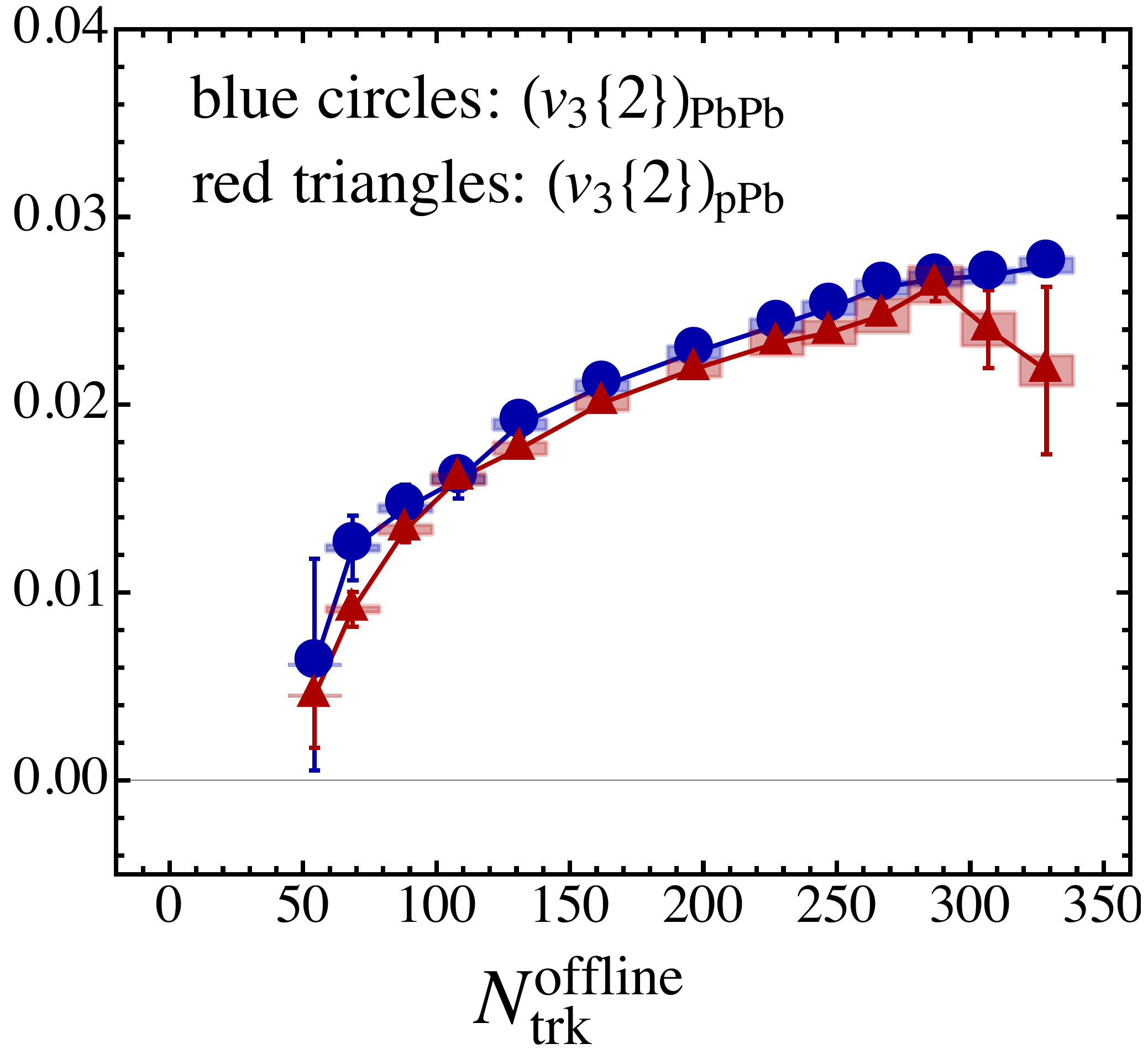}
\caption{The integrated $v_3\{2\}$ for PbPb and pPb vs. multiplicity from
\cite{CMS_flow}. An approximately conformal response leads to $(v_3\{2\})_{\rm pPb}\simeq
(v_3\{2\})_{\rm PbPb}$ at fixed multiplicity.}
\label{fig_int_v32}
\end{figure}

%}~\cite{Bhalerao:2011bp,Jia:2012ju}
% in the independent cluster model\footnote{
%   This equations differs from eq.~(15) of \Ref{Jia:2012ju} beceause
%   we have adopted the $r^3$ weight in \Eq{r3weight}  rather than an $r^2$ weight of \Ref{Jia:2012ju}. 
%If an $r^2$ weight is used then all fluctuation driven eccentricities are equal~\cite{Jia:2012ju}:
%   \st
%   \llangle \delta \epsilon_n^2   \rrangle = \frac{\llangle r^4 \rrangle}{N_{\rm clust} \llangle r^2 \rrangle^2 } \qquad \mbox{ ($r^2$ weight) }
%   \stp
%}~\cite{Bhalerao:2011bp,Jia:2012ju}
%\begin{equation}
%\langle\delta\epsilon_3^2\rangle={\langle r^6\rangle\over N_{\rm clust}\langle r^2\rangle^3}.
%\label{r3}
%\end{equation}
Enforcing conformal dynamics on the linear response, we are led to the conclusion that the triangular flow in  $p+A$ and $A+A$ collisions at a given multiplicity should be approximately the same,
\begin{align}
(v_3\{2\})_{pA}
=&k_3 \sqrt{\langle\delta\epsilon_3^2\rangle_{pA}} \, ,  \\
(v_3\{2\})_{AA}=&k_3 \sqrt{\langle\delta\epsilon_3^2\rangle_{AA}} \, .
\label{v32_int}
\end{align}
 Again, the linear response coefficient $k_3$ is constant at fixed multiplicity.
In \Fig{fig_int_v32} we compare the CMS measurements of $v_3$ for pPb and PbPb
collisions \cite{CMS_flow}. As in the elliptic case, the agreement
between the $v_3$ measurements is remarkable. Empirically 
the ratio of triangular flows is $(v_3\{2\})_{\rm pPb}/(v_3\{2\})_{\rm
PbPb}\approx0.96$. The deviation from unity could be the result of
 corrections to the conformal scaling, or  to the difference in the
 geometries of the colliding systems. 
 %Indeed the percent between  $\sqrt{\langle \delta\epsilon_3^2\rangle_{pA}}$
 %and $\sqrt{\langle \delta\epsilon_3^2\rangle_{AA}}$ could be larger 
%than in the elliptic case, since there is a higher power of $r$ in
%(\ref{r3}) compared to (\ref{r2}). 
%Nevertheless, it
%seems that  linear response to 
%fluctuation driven eccentricities,  together with
%approximately conformal dynamics, 
%gives a reasonable account of the elliptic and triangular flows in pPb collisions at the LHC.

\subsection{Momentum dependence of the flow coefficients}

Having provided a simple explanation for the integrated flow coefficients,
which captures the essential physics, we now study the momentum
dependence. The conformal scaling that we discussed in \Sect{conformal},
suggests that each dimensionful observable can be written
as  the initial temperature $T_{\rm i}\propto Q_s$ to the appropriate power,  times  a dimensionless function
of  $T_{\rm i} L$. $T_{\rm i} L$ is constant at fixed  multiplicity and is thus independent of
the colliding system.   
In
particular, we expect the mean transverse momentum at fixed multiplicity
to be larger in $p+A$  than in $A+A$, 
since the  $p+A$ system has a smaller transverse size.
The expected increase  in $p+A$ of  the mean $\llangle p_T \rrangle$ 
and radial flow was also pointed out in \cite{shuryak_zahed}, and
has been confirmed by the ALICE collaboration~\cite{ALICE_mean_pt,Abelev:2013haa}.
In addition, a dimensional analysis along these lines was recently used to analyze particle spectra in high multiplicity $p+A$ events~\cite{McLerran:2013oju}.
%This
%implies that the radial flow in $p+A$ collisions is larger. 
%Recently, the ALICE collaboration reported that $\langle p_T\rangle$ is indeed
%larger in pPb collisions than in PbPb \cite{ALICE_mean_pt}. 

In the small momentum regime $p_T \sim \llangle p_T \rrangle$, the flow coefficients grow linearly with momentum. Using the conformal scaling, we expect that
\begin{equation}
   \label{conformal_scale}
   \frac{v_n}{\epsilon_n}=\xi_n {p_T\over\langle p_T\rangle}\, ,  
\end{equation}
where the dimensionless slopes $\xi_n$ depend only on the ratio of mean free path to
system size, and are the same for $p+A$ and $A+A$ at fixed multiplicity. 
 Starting from the observation that $\langle p_T\rangle$
 in pPb is roughly 1.25 times higher than in PbPb~\cite{ALICE_mean_pt,Abelev:2013haa}, we
 will rescale the $p_T$ axes of the momentum dependent flow coefficients
 $(v_2\{2\}(p_T))_{\rm PbPb, rscl}$ and $(v_3\{2\}(p_T))_{\rm PbPb}$ with the factor
\begin{eqnarray}
   \label{kappa_def}
 \kappa\equiv{\langle p_T\rangle_{\rm pPb}\over\langle p_T\rangle_{\rm PbPb}}\approx1.25,
\end{eqnarray}
to  compare the dimensionless slopes in the two colliding systems.
 Thus, for  $v_2$ we expect  the following scaling relation between
 the pPb and PbPb systems
 \st
 \left(v_2\{2\}(p_T)\right)_{\rm pPb} = \sqrt{1 -  \frac{\epsilon_s^2}{\epsilon_{2}\{2\}^2 }} \left(v_2\{2\}\left(p_T/\kappa\right)\right)_{\rm PbPb}   \,  .  
 \stp

\begin{figure}
\includegraphics[width=\textwidth]{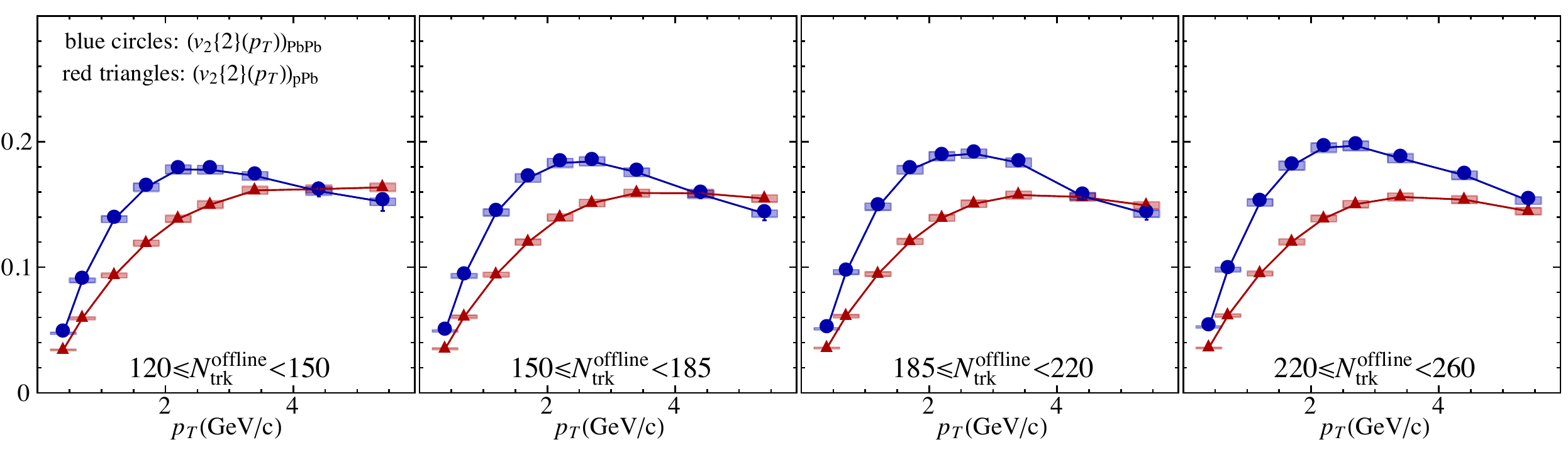}
\includegraphics[width=\textwidth]{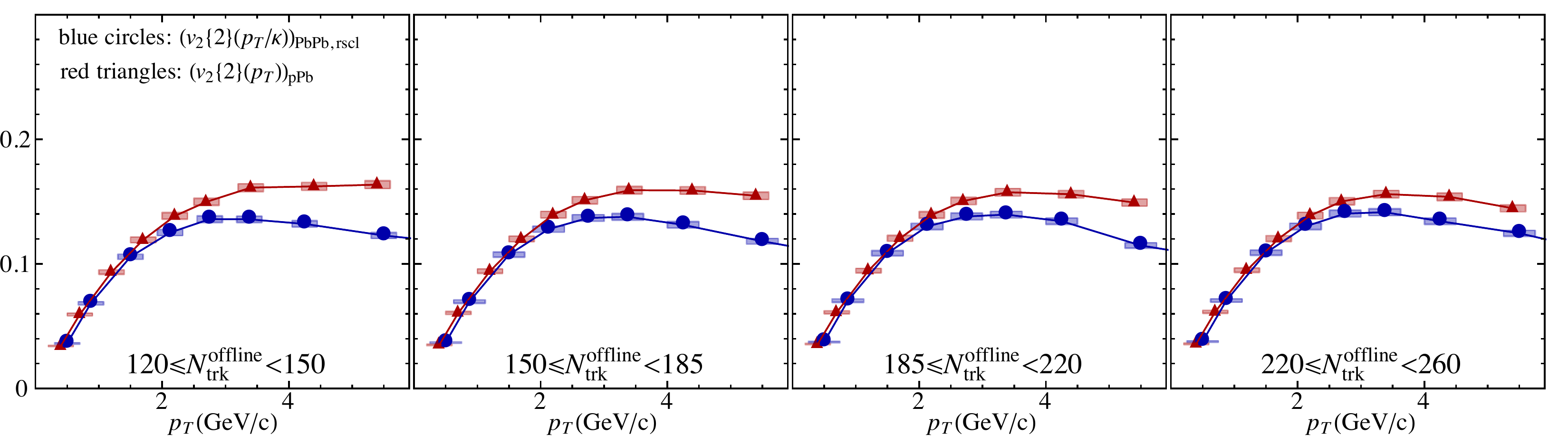}
\caption{A comparison of the momentum dependent $v_2\{2\}$ in pPb and PbPb
collisions. Upper: Original data. Lower: The PbPb data is rescaled to isolate
the fluctuation driven part of the elliptic flow as defined in \Eq{v2_scaling}.
The momentum axis is also scaled by the conformal scaling factor
$\kappa\approx1.25$, \Eq{kappa_def}. 
This is a parameter free rescaling.
The agreement in the low $p_T$ region suggests
that elliptic flow in $p+A$ results from a linear response to the
fluctuations of the initial geometry 
which is conformally  related  to the $A+A$ response.
The data are from \Ref{CMS_flow}.   
}
\label{fig_v22}
\end{figure}

\begin{figure}
\includegraphics[width=\textwidth]{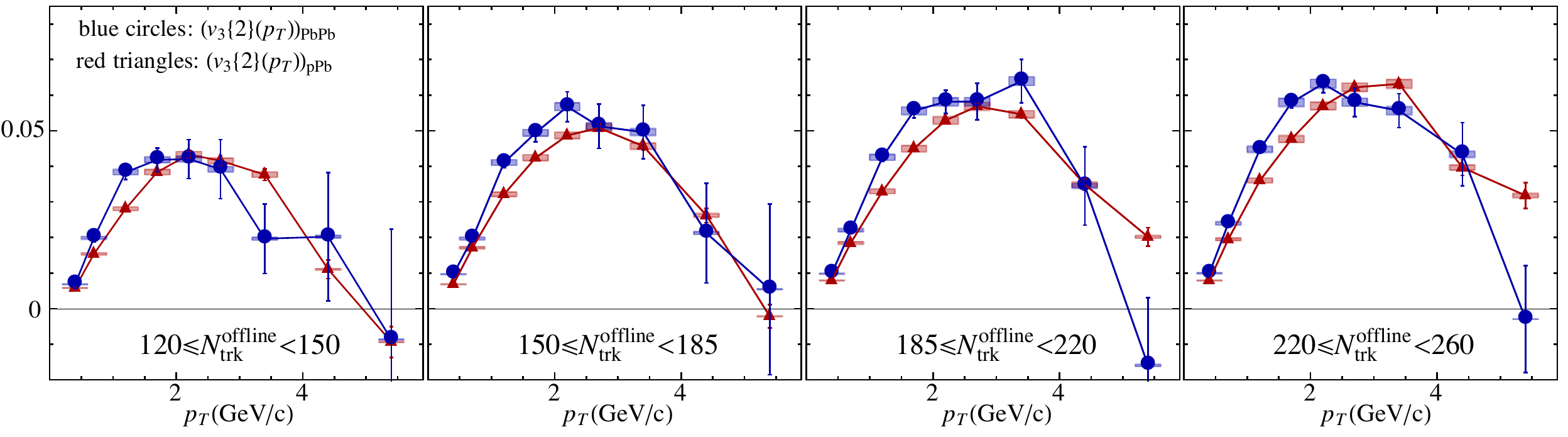}
\includegraphics[width=\textwidth]{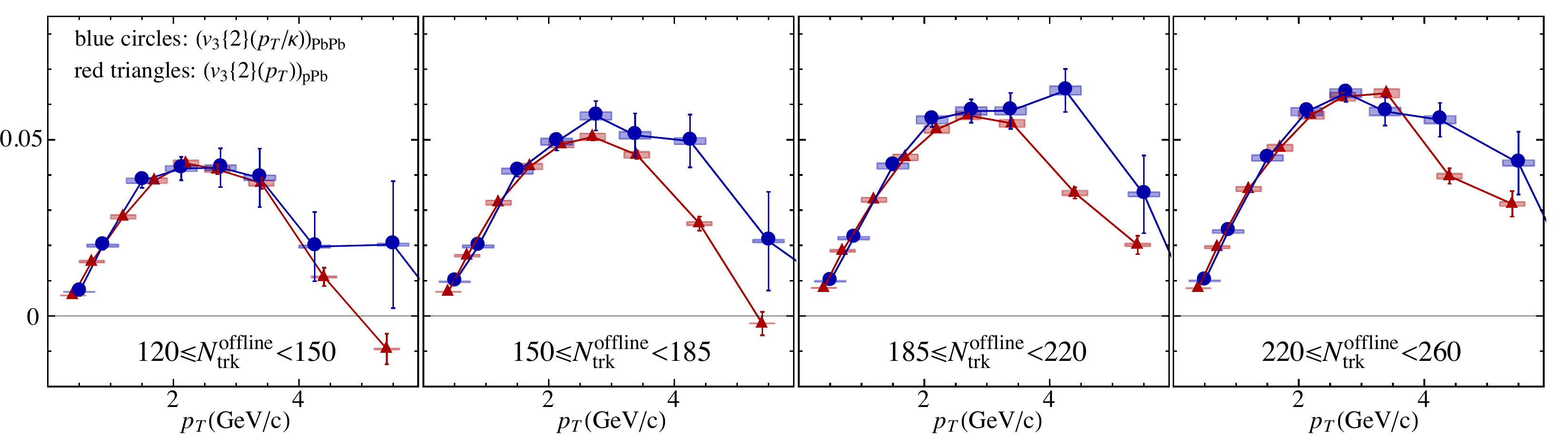}
\caption{The comparison of the momentum dependent $v_3\{2\}$ in pPb and PbPb
collisions. Upper: Original data. Lower: The momentum axis is scaled by the
conformal scaling factor $\kappa\approx1.25$, \Eq{kappa_def}.  This is a parameter free rescaling.
The  agreement in the
low $p_T$ region suggests that the triangular flow in $p+A$ results from a
linear response to the fluctuations in the initial geometry  which 
is conformally related to the $A+A$ response.
The data are from \Ref{CMS_flow}.   }

\label{fig_v32}
\end{figure}

 The original data for $v_2$ and $v_3$
 together with this complete (and parameter free) rescaling is shown in \Fig{fig_v22} and \Fig{fig_v32}  respectively.
 From the
 lower panels in \Fig{fig_v22} and \Fig{fig_v32}, we see that the
 agreement between the dimensionless slopes in the low $p_T$ region is remarkable, and 
 seems to affirm the conformal rescaling. At higher $p_T$, the $v_2\{2\}$
 start to systematically differ. This difference seems to become larger for lower
 multiplicities where  non-flow  could become significant.
 
% We use the flow
% measurements of the CMS collaboration \cite{CMS_flow}. We would like to remind
% the reader that for $v_2$, we are comparing the fluctuation driven part of
% $(v_2\{2\})_{PbPb}$, which differs by the nontrivial correction factor
% (\ref{v2_scaling}) from the original value, with $(v_2\{2\})_{pPb}$. 

An
immediate consequence of the conformal scaling in \Eq{conformal_scale} is that
the
breakdown of the linear regime, where the flow coefficients peak and start to
decrease for larger $p_T$, should happen at a larger $p_T$ for $p+A$ compared to $A+A$.
By comparing the pPb and PbPb measurements in the upper panels in
\Fig{fig_v22} and \Fig{fig_v32} we can see that the maximum for both
$v_2\{2\}$ and  $v_3\{2\}$ is systematically at larger $p_T$ in pPb. 
Rescaling the $p_T$ axis by $\kappa$ as motivated by 
the conformal scaling brings these maxima into alignment.
%In summary, the $p_T$ dependence of the elliptic and triangular 
%flow coefficients seem to corroborate the conformal scaling advocated
%in \Sect{conformal}.
%The
%falloff in $v_2\{2\}$ is not very explicit as opposed to $v_3\{2\}$, which is
%perhaps due to the non-flow effects.

%
% CHANGE
%
It would be interesting to extend this analysis to different particle
species.  We are assuming that the fully inclusive $v_2(p_T)$ best reflects
the conformal dynamics of the initial state.  At freezeout, the dyanmics can
not be strictly conformal \cite{Niemi:2014wta}, and the presence of additional
scales means that different particle species can receive different viscous
corrections \cite{Dusling:2009df}. 

%There is an alternative way to understand the shift of the maximum of
%$v_n(p_T)$ in $p_T$ as one goes from $A+A$ to $p+A$ collisions. In $A+A$ collisions, the linear low
%$p_T$ regime of $v_n(p_T)$ can be  controllably described by  viscous
%hydrodynamics. The particle spectra are computed using equilibrium distribution
%function   and its first viscous correction, $f^{(0)} + \delta f$.
%%The
%%particle spectra is then calculated using Cooper-Frye formula. 
%The hydrodynamic expansion 
%breaks down when  the first viscous correction to the
%distribution,  $\delta f$, become comparable to the
%equilibrium distribution, $f^{(0)}$. The $p_T$ where this breakdown occurs can be estimated by
%noting that  \cite{DMT}
%\begin{equation}
%   \label{df}
%   {\delta f\over f^{(0)}}\sim \left({p\over T}\right)^\alpha {\eta\over s} {1\over T L}
%\end{equation} 
%where the dependence on the system size comes from the derivative expansion. 
%%As
%%discussed above, the conformal scaling implies that $\eta/s \prot$, which
%%is fulfilled in weakly coupled QCD and also clearly in the strong coupling
%%results from AdS/CFT. 
%Further assuming that the relevant temperature scale in  \Eq{df}
%scales with the initial temperature, we infer that the
%breakdown of the linear regime in $v_n(p_T)$ is proportional to the 
%initial temperature at fixed multiplicity $p_T \propto T_{\rm i}$, and therefore
%scales by the constant factor $\kappa$ from $A+A$ to $p+A$ collisions.
%%\textcolor{blue}{One might be worried about the effects of the freezout
%%spoiling the conformal scaling argument. However, .....}

\section{A qualitative energy loss analysis for $p+A$ }
\label{jets}

In this section we will qualitatively sketch the implications of the conformal scaling discussed in
\Sect{conformal} for parton energy loss. 
For reviews of energy loss see \cite{Baier:2000mf,Arnold:2009mr,Majumder:2010qh}.  
A hard parton of energy $E$, traveling in the medium experiences energy loss
from mainly two sources: collisions in the medium and medium induced radiation.
  The collisional energy loss can be parametrized by the drag
coefficient $\hat e$, where  $\dd E/\dd t= -
\hat e$. To estimate the medium induced radiation, we will adopt the BDMPS framework~\cite{BDMPS,Baier:1996kr},  giving a  heuristic review  before discussing the
implications for  $p+A$ collisions \cite{[{A very useful guide
         to the current discussion and relevant literature is:
}] Arnold:2009mr}. 

The underlying physics can be understood as the interplay among the different
scales in the problem: the formation length $\ell_{\rm form}\sim{\omega/
k_\perp^2}$, the mean free path $\ell_{\rm mfp }$, and the system size $L$. The
accumulation of transverse momentum squared $\llangle k_\perp^2 \rrangle $ of the radiated gluons
as the parton traverses the medium is modeled by a random walk in
momentum space with  diffusion coefficient $\hat q$, where  $ \hat
q=\dd\llangle k_\perp^2 \rrangle/\dd t$. 
The medium induced radiation
spectrum has several regimes, 
depending on the frequency $\omega$ of the radiated gluon: 
\begin{enumerate}[(i)]

   \item In the Bethe-Heitler regime  where  $\omega<\hat q \,   \ell_{\rm mfp }^2$ and  $\ell_{\rm form }<\ell_{\rm mfp }$, the radiation spectrum 
 is of order
 \begin{align}
    \omega {\dd N_g \over \dd \omega \,\dd z}\sim & {\alphas \over \ell_{\rm mfp}}  \qquad \qquad (\omega<\hat q\, \ell_{\rm mfp}^2) \,.   \tag{Bethe-Heitler} 
\end{align}
The radiation in this soft frequency range can be neglected in simulations
of parton energy loss.

\item  In the LPM regime where  $\hat q \,\ell_{\rm mfp}^2 <\omega <\hat q\, L^2$  and  $\ell_{\rm mfp}<\ell_{\rm form }<L$, the radiation is depleted by destructive interference
between several subsequent scatterings.  Effectively $N=\ell_{\rm form}/\ell_{\rm mfp}$ scatterings act like one scattering center for
the induced radiation. This is the \LPM\,(LPM) effect, and the formation time in this
regime  should be calculated self consistently in order to take into account of
the destructive interference. Since the average $k_{\perp}^2$ after $N$
collisions is $\hat q\,\ell_{\rm form}$, we obtain the relation $\ell_{\rm form}\sim{\omega / \hat q\,\ell_{\rm form}}$. Thus, the gluon spectrum in the LPM regime is   of order
\begin{alignat}{2}
\label{LPM}
\omega {\dd N_g \over \dd \omega \;\!\dd z}\sim & {\alphas \over \ell_{\rm mfp}}{1\over N}\sim{\alphas \sqrt{\hat q \over \omega} }  
\qquad \qquad  & (\mbox{with } \hat q\, \ell_{\rm mfp}^2<\omega<\hat q\, L^2) \, .  
\tag{LPM} 
\end{alignat}

\item  Finally in the deep LPM regime where  $\omega>\hat q\,L^2$,
the formation length of the radiation  exceeds the size of the medium 
$\ell_{\rm form} > L$,  and the medium acts as a single scattering center.
In this regime the medium induced radiation
 spectrum  is  of order
 \begin{alignat}{2}
     \omega \frac{\dd(\Delta N_g)}{\dd\omega}   \sim& \alphas \left(\frac{L}{\ell_{\rm form}} \right)^2 \sim  \alphas \frac{(\hat q L^2)^2}{\omega^2}  
     \qquad\qquad & (\mbox{with } \omega>\hat q\, L^2) \, ,  
    \tag{deep-LPM} 
    \label{deep-LPM}
\end{alignat}
where $\Delta N_g = N_g - N_g^{\rm vac}$ is the number of 
gluons emitted in excess of the vacuum shower.
\end{enumerate}

The relation between the average energy loss $\Delta E$ and 
the system size
depends on 
the initial energy of the parton. For example, for $E< \hat q \,L^2$ the parton
never experiences the deep LPM regime.
In this case, the average energy loss is found
by integrating the appropriate 
radiation spectrum (\ref{LPM}) over
the path length and frequency  from
$\omega=0\ldots E$:  
\begin{alignat}{2}
   \Delta E \sim&\alphas\,\sqrt{E \hat q \phantom{|} } L  \qquad \qquad  &(\mbox{for } E<\hat q\,L^2) \, .
\end{alignat}
A more energetic parton, with $E>\hat q\,L^2$, experiences the deep LPM suppression, and  integrating the 
corresponding radiation spectrum (\ref{deep-LPM})  from
$\omega = \hat q L^2 \ldots \infty$ yields~\cite{BDMPS,Baier:1996kr,Arnold:2009mr} 
\begin{alignat}{2}
   \Delta E\sim& \alphas \,\hat q L^2  \qquad \qquad  &(\mbox{for } E>\hat q \,L^2)\, .
\end{alignat}

%The total radiated energy  in the LPM regime  is given
%\st
%\Delta E= \int_0^{\sim E} \omega  {d N_g \over d \omega dz} \, d\omega dz }  \sim \alphas\,\sqrt{E \hat q} \; L \,, \qquad  E<\hat q\,L^2 \, ,
%\stp
%Similarly 
%\st
% \Delta 
%\stp
%or
%\st
%\Delta E  \sim  \alphas\,\sqrt{E \hat q} \; L \,, \qquad  E<\hat q\,L^2 \, .
%\stp
%The relation between the total energy loss $\Delta E$ and the system size depends on the initial energy of the parton. For $E< \hat q \,L^2$, the parton never experiences the deep LPM regime and the total energy loss is given by
%\bg
%\Delta E\sim\alphas\,\sqrt{E \hat q} \,L\quad,\quad E<\hat q\,L^2.
%\nd 
%However a more energetic parton with $E>\hat q\,L^2$ experiences the deep LPM suppression and the total energy loss is~\cite{BDMPS}
%\bg
%\Delta E\sim \alphas \hat q\, L^2 \quad E>\hat q \,L^2.
%\nd
%The leading parton energy energy loss due to induced radiation in this 
%frequency range is
%\st
%\Delta E=\int_0^{\sim E} \omega \frac{dN_g}{d\omega dz} \, d\omega dz \,  .
%\stp
%or
%\bg
%\Delta E\sim\alphas\,\sqrt{E \hat q} \,L\quad,\quad E<\hat q\,L^2.
%\nd 

We can now discuss the implications of the conformal scaling framework for jet
energy loss. Let us denote the critical energy that separates these two regimes
as $E_{\rm cr}= \hat q \, L^2$. The conformal scaling from $A+A$ to $p+A$ predicts
that $E_{{\rm cr},pA}= \hat q_{pA} \,L_{pA}^2= \kappa\, \hat q_{AA} \,L_{AA}^2$ where
$\kappa=L_{AA}/L_{pA}$ is the scaling factor. This scaling of $E_{cr}$ from
$p+A$ to $A+A$ follows from $\hat q\sim T^3$ and the prediction of the
conformal dynamics where $T_{pA}=\kappa T_{AA}$. Since $E_{{\rm cr},pA}>E_{{\rm cr},AA}$,  the deep LPM regime (which is associated with
\textit{small} systems) is achieved \textit{later} as a function of increasing
total parton energy $E$ for the $p+A$ collisions. This  
counter-intuitive result occurs because, in addition to the decrease in
the system size, the conformal scaling leads to an increase  in 
$\hat{q}$.
%in the average
%transverse momentum accumulated.
The increase in $\hat q$ translates into a decrease in the typical formation
length, requiring more energy to reach the transitional point where the formation
length exceeds the system size. The same reasoning also predicts somewhat larger
transverse momentum broadening for jets produced in $p+A$ collisions.
%compared
%to $A+A$ collisions. 

A more quantitative analysis of jet energy loss in $p+A$
 is left for future work.  We hope that qualitative 
(and counter-intuitive) features of the conformal scaling
outlined in this section can survive 
in a more complete treatment  of parton energy loss.

\section{Summary and Discussion}
By analyzing the flow measurements of pPb and PbPb collisions at the LHC with several physically motivated rescalings, we provide evidence for  a collective response to the geometry in high multiplicity pPb collisions.

First, we note  that once the average ellipticity is scaled out of the PbPb elliptic flow, the
fluctuation driven integrated $v_2\{2\}$ in PbPb is the same as in
pPb at fixed multiplicity (\Fig{fig_int_v22}). The integrated triangular flows in these two colliding systems are already equal.  It seems to us
phenomenologically untenable to ascribe different physics  to the $p+A$ and
$A+A$ flow measurements. Since the rescaling in PbPb was entirely motivated by linear response and geometry,
we conclude that both the elliptic and triangular flow in pPb should also be
understood as a linear response to initial geometric fluctuations.  
\Sect{conformal} and \Sect{eccentricity} offer a direct explanation for why the response coefficients and fluctuation driven eccentricities in these two systems are similar at fixed  multiplicity.

First, a simple estimate based on approximate conformal symmetry at high energies shows that the
mean free path to system size in the two systems is constant at fixed multiplicity (see \Sect{conformal}).
Thus, the dynamical response of the $p+A$ and $A+A$ systems are related by a simple
conformal rescaling of the initial 
temperature and the system size such that $\ell_{\rm mfp}/ L \propto 1/(T_{\rm i}
L)=\mbox{constant}$.  The ${\rm pPb}$ system is smaller, but also hotter, leading to
the same response at fixed multiplicity.

Next we used the 
independent cluster model to estimate the eccentricities in both 
systems. (In PbPb the independent cluster model reproduces
the results of more sophisticated Glauber models~\cite{Bhalerao:2011bp}.)
Assuming that the multiplicity is proportional to the number of clusters,
we find that the ratio of fluctuation-driven eccentricities in the two colliding systems is determined by a square root of a geometric double ratio,
{\it e.g} 
\st
\sqrt{ \frac{\langle\delta\epsilon_2^2\rangle_{pA}}{\langle\delta\epsilon_2^2\rangle_{AA} } } = 
 \sqrt{ \frac{(\llangle r^4 \rrangle/\llangle r^2 \rrangle^2)_{pA} }{(\llangle r^4 \rrangle/\llangle r^2 \rrangle^2)_{AA} } } \, .
\stp
The importance of this and related formulas is that even quite different $p+A$ profiles lead to
approximately the same $\sqrt{\llangle \delta \epsilon_2^2 \rrangle }$ and $\sqrt{\llangle \delta \epsilon_3^2\rrangle }$.
%Thus the fluctuations driven
%eccentricities  in the two colliding systems differ by at most $10\%$ at
%fixed multiplicity. To re-iterate,
Without fine tuning the profile it is reasonable to expect that the fluctuation-driven eccentricities in the two systems are equal  to $\sim5\%$ accuracy. For a Gaussian $p+A$ profile, which arises in any diffusive process and seems particularly apropos, this double ratio is shown in \Fig{eratio_fig} and is close to unity for both the second and third eccentricities.

The $p_T$ dependence of the elliptic and triangular flow gives additional
evidence supporting the conformal scaling described above. The $\langle p_T\rangle$ and the slope of both $v_2\{2\}(p_T)$, $v_3\{2\}(p_T)$ scale in the same way between pPb and PbPb as expected from the conformal scaling of \Sect{conformal}.  
Indeed, the rescalings in \Fig{fig_v22} and \Fig{fig_v32} are essentially
parameter free, given the measured $\llangle p_T \rrangle$ in both colliding
systems.  The agreement between the dimensionless slopes in the low $p_T$
region 
in these figures corroborates the conformal scaling  outlined in \Sect{conformal}.

%
% CHANGE
%
Finally, we have outlined several qualitative expectations of conformal scaling for energy loss.  In particular, 
the  finite size transition in energy loss, from  a linear ($\Delta E \propto L$) to a quadratic $(\Delta E \propto  L^2$) length dependence,
requires \emph{higher} energy for the initial parton in the $p+A$ system.
While a quantitative discussion and simulation of energy loss is left for
future work, the conformal scaling arguments of \Sect{jets} suggest that the
energy loss and transverse momentum broadening of jets should be somewhat
larger in $p+A$ than in  $A+A$  at the same multiplicity.  Since the energy loss
in $A+A$ is fairly mild in these peripheral bins\footnote{See for example the 50-60\% centrality bin in Fig. 6 of \Ref{Aad:2012vca}.}, and since preliminary
measurements of jet energy loss in $p+A$ are at rather low multiplicity
\cite{CMS:2014qca}, this prediction does not seem in
contradiction with current measurements, which do not indicate energy loss.

In summary, we have provided a concise explanation for why the angular correlations 
in pPb and PbPb collisions are similar -- these correlations are 
the result of an approximately conformal response to 
fluctuation-driven eccentricities. 
It is important to 
emphasize that any conformal response
to the geometry will yield similar correlations in the two colliding  systems.
However, it is equally important to emphasize that any conformal dynamics
will asymptote to conformal hydrodynamics in the limit of high multiplicity.

\section{Acknowlegments}
We thank A. Bzdak, D. Kharzeev, R. Venugopalan, E. Shuryak, L. Yan, Y. Yin, and
I. Zahed for useful discussions. We also thank J.-Y. Ollitrault, U. Heinz, and T. Schaefer for a discussion of the $\dd N/\dd y$ scaling of viscous corrections. This work was supported by the U.S. Department
of Energy under the grants DE-FG-88ER40388 (GB) and DE-FG-02-08ER4154 (DT).

%\subsection{A quantitative energy loss analysis}
%
%To simulate the energy loss of jets, we follow a model
%due to Muller and Guangyou which implements the following
%model. 
%
%First we generate partons using Pythia 8. The 

%\bibliographystyle{apsc4-1}
\bibliography{pAbib} 

\end{document}